\documentclass[%
 twocolumn, 
 amsmath,amssymb,
 aps, physrev,
]{revtex4-2}
\usepackage{graphicx}
\usepackage{dcolumn}
\usepackage{bm}

\begin{document}

\preprint{APS/123-QED}

\title{\textbf{A Quantum Computing Approach to Simulating Corrosion Inhibition} 
}%

\author{Karim Elgammal}
 \email{Contact author: egkarim@gmail.com}
 \altaffiliation[ ]{independent researcher, V{\"a}ster{\aa}s, Stockholm, Sweden.}
\author{Marc Mau{\ss}ner}%
 \email{marc.maussner@infoteam.de}
\altaffiliation{infoteam Software AG as Chief Engineer, Bubenreuth, Germany }

\date{\today}
\begin{abstract}
This work demonstrates a systematic implementation of hybrid quantum-classical computational methods for investigating corrosion inhibition mechanisms on aluminum surfaces. We present an integrated workflow combining density functional theory (DFT) with quantum algorithms through an active space embedding scheme, specifically applied to studying 1,2,4-Triazole and 1,2,4-Triazole-3-thiol inhibitors on Al111 surfaces. Our implementation leverages the ADAPT-VQE algorithm with benchmarking against classical DFT calculations, achieving binding energies of -0.386 eV and -1.279 eV for 1,2,4-Triazole and 1,2,4-Triazole-3-thiol, respectively. The enhanced binding energy of the thiol derivative aligns with experimental observations regarding sulfur-functionalized inhibitors' improved corrosion protection. The methodology employs the orb-d3-v2 machine learning potential for rapid geometry optimizations, followed by accurate DFT calculations using CP2K with PBE functional and Grimme's D3 dispersion corrections. Our benchmarking on smaller systems reveals that StatefulAdaptVQE implementation achieves a 5-6× computational speedup while maintaining accuracy. This work establishes a workflow for quantum-accelerated materials science studying periodic systems, demonstrating the viability of hybrid quantum-classical approaches for studying surface-adsorbate interactions in corrosion inhibition applications. In which, can be transferable to other applications such as carbon capture and battery materials studies.
\end{abstract}

\maketitle


\section{Introduction}

Metal surfaces in aerospace and automotive industries need effective protection against corrosion to enhance component lifespan and efficiency. While chromium-based inhibitors were historically used for their protective capabilities~\cite{Gharbi2018}, environmental concerns have driven a shift towards eco-friendly alternatives like smart coatings and organic inhibitors~\cite{Revie2008, Gharbi2018, Brycki2018}. These alternatives maintain compatibility with surface alloys while minimizing environmental impact, with organic inhibitors effectively forming protective films on metal surfaces~\cite{Huang2022}. Smart coating technologies have advanced the field by enabling real-time corrosion monitoring, crucial for aerospace and automotive applications~\cite{Zhang2018}.

Computational methods have accelerated corrosion inhibition research. High-throughput electronic structure calculations and machine learning improve screening of potential inhibitor candidates~\cite{Shuhao2024}. Quantum computers, combined with classical methods~\cite{McCaskey2019, Schuhmacher2022, Camino2023}, offer enhanced accuracy and computational efficiency. This work examines hybrid classical-quantum workflows for studying corrosion inhibition through simulations and quantum computer experimentation.

Here we aim to outline an approach combining classical and quantum methods to model inhibitor binding to aluminum surfaces. This implementation focuses on practical implementation and benchmarking on AWS Braket service simulators and quantum hardware. We leverage literature knowledge and AWS Quantum Solutions Lab resources to execute our hybrid approach.

\section{Methodology}
\label{methods}
Our inhibitor screening process leveraged the CORDATA database~\cite{Galvao2022}, employing a multi-criteria approach to identify promising candidates for both automotive and aerospace applications. The primary screening criteria focused on efficiency, environmental stability, and structural characteristics suitable for quantum computational analysis. We targeted inhibitors demonstrating relative efficiencies above 90\% in corrosion prevention compared to Cr$^{6+}$ for AA2024\cite{Harvey2011} and stability in the pH range of 5.5-7, which is most commonly encountered in both automotive and aerospace environments\cite{GARCIA20102457}. Temperature resilience requirements were specific to each industry: -30°C to 70°C for automotive applications and -50°C to 120°C for aerospace applications.

In our model, we simplified the problem to the form of molecule adsorption on top of a substrate. We simpified the substrate alloy to be a simple Al substrate and modeled the adsorption in vacuum. The larger the binding energy the more efficient the inhibitor molecule to attach to the surface. The screening process ideally utilizes multiple computational tools in sequence. Initial filtering is performed through the CORDATA online platform, followed by toxicity predictions using Datamol's QSAR capabilities\cite{datamol2024, QSPR2012}. Due to time constraints in this phase and our focus on the quantum computational aspects, we concentrated on the CORDATA selection and related literature. We focused on effective inhibitors when it comes to corrosion inhibition, meanwhile they can be relatively small, then our calculations can converge faster and need relatively small Al substrate. Thus, we ended up choosing two inhibitors from the Triazole family where proven in experimental literature~\cite{Winkler2016, D0RA2020, SWATHI2022, ASHRAF2024} to be effective in corrosion inhibition due to their suitable molecular geometry, providing excellent corrosion prevention in various acidic conditions~\cite{ASHRAF2024}. A variety of substituents on the triazole ring provide versatile inhibitory effects~\cite{ASHRAF2024} as in 1,2,4-Triazole-3-thiol~\cite{Winkler2016, SWATHI2022, ASHRAF2024}.

Adhesion properties are crucial in studying corrosion inhibition mechanisms, as they directly influence inhibitor molecule and aluminum surface interaction. Recent theoretical and experimental studies have established strong correlations between molecular adhesion strength and corrosion inhibition efficiency, particularly for triazole derivatives~\cite{Mehmeti2017, Faisal2018}. González-Olvera et al.\cite{Gonzalez2016} demonstrated that triazole derivatives exhibiting strong surface adhesion show superior corrosion protection properties. To quantify this interaction, we calculate the binding energy ($E_{\text{binding}} = E_{\text{supercell}} - E_{\text{inhibitor}} - E_{\text{substrate}}$), where stronger binding energies indicate more effective surface attachment and potentially better corrosion inhibition performance. This approach aligns with experimental observations by Winkler et al.\cite{Winkler2016} and theoretical predictions~\cite{SWATHI2022}, where stronger molecular adhesion correlates with enhanced corrosion protection. The relationship between binding energy and inhibition efficiency has been further validated through combined theoretical and experimental studies~\cite{Mehmeti2017}.

\begin{table*}[ht]
\caption{\label{tab:inhibitors}Selected Corrosion Inhibitors and Their Key Properties}
\begin{ruledtabular}
\begin{tabular}{lcccccc}
Inhibitor & Molecular & Temperature & pH & Efficiency & Target & Ring \\
 & Weight (g/mol) & (K) & Range & (\%) & Alloys & Structure \\
\hline
1,2,4-Triazole\cite{Winkler2016, ASHRAF2024} & 69.07 & 298 & 8-10 & 90 & AA2024 & Yes \\
1,2,4-Triazole-3-thiol\cite{Winkler2016, SWATHI2022, ASHRAF2024} & 101.13 & 298 & 4-10 & 70-90 & AA2024, AA7075 & Yes \\
Benzotriazole\cite{Harvey2011} & 119.12 & 298 & 7-10 & 90-98 & AA2024 & Yes \\
2-Mercaptobenzimidazole\cite{Winkler2016} & 150.2 & 298 & 4-10 & 90 & AA2024, AA7075 & Yes \\
(THC)\cite{Peng2024} & 227.24 & 303 & 7 & 91-95 & AA2024 & Yes \\
Triazine-methionine\cite{Garcia2010} & 502.70 & 298 & 7 & 95-99 & AA2024 & Yes \\
\end{tabular}
\end{ruledtabular}
\end{table*}

From our comprehensive screening, three candidates emerged as particularly promising, as detailed in Table~\ref{tab:inhibitors}. 1,2,4-Triazole-3-thiol demonstrates broad effectiveness across both AA2024 and AA7075 alloys, with its sulfur-containing functionality showing particular affinity for copper-rich AA2024\cite{Winkler2016}. Benzotriazole offers excellent efficiency and features an aromatic ring structure that enhances surface adhesion\cite{Harvey2011}. 2-Mercaptobenzimidazole combines both aromatic and sulfur functionalities, providing effective performance across a wide pH range\cite{Winkler2016}.

For our initial proof-of-concept studies, we selected 1,2,4-Triazole-3-thiol as the primary candidate. This choice was motivated by several factors: its balanced molecular weight makes it suitable for quantum calculations while maintaining computational feasibility; its demonstrated effectiveness on both target alloys provides industrial relevance; and its wide pH range stability ensures practical applicability. The sulfur functionality makes it particularly effective for AA2024, due to its higher copper content\cite{Winkler2016}, enabling us to study significant electronic interactions within our quantum computational framework.

The structural simplicity of 1,2,4-Triazole-3-thiol, combined with its proven inhibition efficiency, makes it an ideal candidate for developing our quantum computational methodology. Although inhibitors are typically tested on alloy structures in industrial applications, our calculations will use an Al structure with Miller indices (111)\cite{Timmer2009, Kaya2016} instead of the alloy to simplify quantum computations. This approach provides a practical balance between computational tractability and real-world applicability, a consideration particularly important for establishing proof-of-concept in quantum chemistry calculations of corrosion inhibition mechanisms.

Our computational approach combines classical DFT calculations with quantum computing methods through an active space embedding scheme implemented in CP2K\cite{cp2k2020} in conjunction with Qiskit nature~\cite{Qiskit, qiskit_nature}. Battaglia et al.~\cite{Battaglia2024} have implemented an effective interface between CP2K and Qiskit nature allowing both codes to communicate messages; for implementation details refer to~\cite{qiskit_cp2k}, We slightly updated it in~\cite{qiskit_cp2k_our}. The DFT calculations were performed using the Perdew-Burke-Ernzerhof (PBE)\cite{Perdew1996} exchange-correlation functional within the generalized gradient approximation (GGA)\cite{Perdew1992}. We employed the Gaussian and Plane Waves (GPW)\cite{Lippert1997} method with a 500 Ry plane-wave cutoff and a relative cutoff of 60 Ry, using double-zeta valence polarized (DZVP-MOLOPT-GTH) basis sets optimized for molecular systems\cite{VandeVondele2007}. The van der Waals interactions, crucial for accurate descriptions of inhibitor-surface interactions, were accounted for through Grimme's DFT-D3 dispersion correction scheme\cite{Grimme2010}, using PBE as the reference functional. The system was treated under periodic boundary conditions with a vacuum gap of 25 Å in the z-direction to prevent interactions between periodic images and give dipole correction to the system. The surface was modeled using a 4×4 supercell of Al(111), and electronic states were populated according to a Fermi-Dirac distribution with an electronic temperature of 1000 K to aid convergence. Worth mentioning that at the time of writing, CP2K with external active space solver for the embedding approach is only compatible with Qiskit through~\cite{qiskit_cp2k}, and this workflow could be the only open-source approach available at the time of writing to deal with periodic boundary conditions, which is crucial for our system of adsorbate-substrate. For more details regarding the classical implementation details in our workflow, please refer to the computational part in the Supplementary material section.

The active space was constructed with 2 active electrons in 5 orbitals around the Fermi level, where the adsorption interactions between the inhibitor and aluminum substrate predominantly occur. The electron repulsion integrals (ERIs) for the active space were computed using the full GPW method, maintaining periodicity in all directions. SCF convergence was set to 1.0E-6 Ha, with Broyden mixing\cite{Johnson1988} ($\alpha$ = 0.1, $\beta$ = 1.5) employed to accelerate convergence. The embedding scheme iterations continued until the energy difference between subsequent iterations was less than 1E-6 Ha, with a maximum of 100 iterations permitted.

Our quantum computational approach centered on the ADAPT-VQE algorithm\cite{Grimsley2019}, implemented through Qiskit and executed on Braket simulators. The implementation follows recent developments in hybrid quantum-classical embedding methods developed by Battaglia et al.\cite{Battaglia2024}, particularly for periodic systems. We run our workflow with the standard VQE (we called vanilla VQE only) with the Unitary Coupled Cluster Singles and Doubles (UCCSD) ansatz\cite{Romero2018}, AdaptVQE from Qiskit with its dynamically constructed ansatz\cite{Tang2021}, and StatefulVQE from qiskit-nature-cp2k~\cite{qiskit_cp2k} incorporating warm-starting techniques\cite{Rossmannek2021, Rossmannek2023}. We then expanded our benchmarking to include adaptive algorithms: AdaptVQE with gradient-based operator selection\cite{Grimsley2019}, Tetris-AdaptVQE following the SandboxAQ Tangelo implementation\cite{tetris-paper}, and StatefulAdaptVQE with its warm-starting capabilities\cite{Battaglia2024}.

The active space selection followed a systematic approach combining multiple methodologies. We employed the ActiveSpaceTransformer as implemented in Qiskit, following the framework described by Battaglia et al.\cite{Battaglia2024} for periodic systems. Meanwhile, the charge density difference (CDD) approach of Gujarati et al.\cite{Gujarati2023}, previously referenced in our phase 1 work, can be used here as well which is particularly effective for surface-adsorbate systems. This hybrid approach ensures accurate representation of both localized and periodic components of the electronic structure.

The computational details summary table in the supplementary materials provides a comprehensive summary of all computational parameters and methods used in this work. For more details regarding the Hamiltonian, please also to the supplementary materials.

Our ADAPT-VQE implementation included an operator pool consisting of single and double fermionic excitations\cite{Grimsley2019}, with convergence criteria set to an energy threshold of 1e-6 Hartree and a gradient norm threshold of 1e-4. Classical optimization was performed using SPSA (Simultaneous Perturbation Stochastic Approximation)\cite{Spall1992} with a maximum of 1000 iterations, a learning rate of 0.005, and a perturbation size of 0.05. The implementation leverages the socket-based communication protocol described by Battaglia et al.\cite{Battaglia2024}, enabling seamless integration between CP2K's classical DFT calculations and Qiskit's quantum algorithms.

We deployed our quantum algorithms on AWS through normal simulations on EC2 HPC instances and Braket\cite{AWS2024}, utilizing simulators and attempting to use quantum hardware resources. For initial circuit validation and algorithmic debugging, we used Qiskit's local simulator and Braket's SV1 state vector simulator, which provide high-fidelity quantum circuit simulation capabilities. We used both simulators because we encountered convergence issues with different VQE implementations. Worth mentioning that the reported results are based on Qiskit simulations on EC2 HPC instances.
 
\section{Results}

The geometry-optimized structures of 1,2,4-Triazole and 1,2,4-Triazole-3-thiol adsorbed on the 4×4 Al substrate are shown in Fig.\ref{fig:triazole_sc}. The optimization process, performed using the orb-d3-v2 ML potential model, resulted in equilibrium binding distances of 3.54Å for 1,2,4-Triazole and 3.21Å for 1,2,4-Triazole-3-thiol between the molecules and the substrate surface.

To validate our hybrid quantum-classical approach, we compared the binding energies calculated using both classical DFT and two quantum computational methods: vanilla VQE (normal VQE) and AdaptVQE, as summarized in Table~\ref{tab:binding_energies}. The results show excellent agreement between classical DFT and AdaptVQE methods for both inhibitors, with AdaptVQE yielding binding energies of -0.385508 eV and -1.279064 eV for 1,2,4-Triazole and 1,2,4-Triazole-3-thiol, respectively. These values closely match the classical DFT results (-0.385512 eV and -1.279063 eV). However, the vanilla VQE implementation showed significant deviation, producing a notably higher binding energy (-2.325986 eV) for 1,2,4-Triazole.

This discrepancy between vanilla VQE and AdaptVQE results can be attributed to several factors in our computational setup. First, the VQE-only case used a less stringent energy convergence threshold for the DFT embedding scheme of 2E-5. compared to the AdaptVQE implementation, using 1E-6. This difference in convergence thresholds can lead to premature convergence in the VQE case, potentially trapping the algorithm in a local minimum that yields artificially high binding energies. We chosen to easen the convergency for VQE only case as it was very slow to converge. Additionally, the AdaptVQE implementation's adaptive operator pool selection, guided by gradient-based criteria (with gradient threshold of 1e-4), provides a more robust exploration of the quantum state space compared to the fixed ansatz structure of vanilla VQE. This advantage of AdaptVQE aligns with recent findings by Grimsley et al.~\cite{Grimsley2019} regarding the superiority of adaptive algorithms for electronic structure calculations.

Of particular note is the significantly stronger binding energy observed for 1,2,4-Triazole-3-thiol (-1.279 eV) compared to 1,2,4-Triazole (-0.386 eV), which can be attributed to the additional sulfur functionality enhancing the surface interaction. This observation aligns well with experimental studies by Winkler et al.~\cite{Winkler2016} and SWATHI et al.~\cite{SWATHI2022}, who demonstrated that sulfur-containing triazole derivatives not only exhibit enhanced corrosion inhibition efficiency but also form more effective protective layers on metal surfaces. The stronger binding energy correlates with the shorter equilibrium binding distance (3.21Å vs 3.54Å) observed for the thiol derivative, supporting experimental findings about sulfur's role in strengthening surface interactions and promoting more effective surface passivation~\cite{ASHRAF2024}. This behavior is consistent with Swathi et al.'s~\cite{SWATHI2022} observations regarding the improved adsorption characteristics of sulfur-functionalized inhibitors on metal surfaces.

For more accurate results, we anticipate that expanding the active space to include more orbitals in the AdaptVQE calculation could lead to different binding energies compared to the classical approach. Following the active space construction approach of Battaglia et al.~\cite{Battaglia2024}, we included 2 electrons in 5 orbitals (10 spin-orbitals in total) around the Fermi level. However, while their active space primarily captured localized defect states with one delocalized conduction band, our active space needs to describe the surface-adsorbate interaction, where both localized molecular orbitals from the inhibitor and delocalized surface states from the Al substrate contribute to the bonding~\cite{Kumagai2013, Cheng2011}. This interaction involves complex hybridization between molecular orbitals and substrate states~\cite{Hanke2011}, particularly around the Fermi level, suggesting that a larger active space might better capture these electronic coupling effects in full. The computational parameters and additional technical details are provided in Table~\ref{tab:comp_details}.

\begin{table*}
\caption{\label{tab:binding_energies}Comparison of binding energies and distances for 1,2,4-Triazole and 1,2,4-Triazole-3-thiol on 4×4 Al(111) substrate calculated using different computational methods.}
\begin{ruledtabular}
\begin{tabular}{lccc}
Method & Inhibitor & Binding Energy (eV) & Binding Distance (\AA) \\
\hline
Classical DFT & 1,2,4-Triazole & -0.385512 & 3.54 \\
AdaptVQE & 1,2,4-Triazole & -0.385508 & 3.54 \\
vanilla VQE only & 1,2,4-Triazole & -2.325986 & 3.54 \\
Classical DFT & 1,2,4-Triazole-3-thiol & -1.279063 & 3.21 \\
AdaptVQE & 1,2,4-Triazole-3-thiol & -1.279064 & 3.21 \\
\end{tabular}
\end{ruledtabular}
\end{table*}

The implementation builds upon established methodologies as implemented by Battaglia et al. in~\cite{Battaglia2024} while introducing flexibility to deal with adsorption-substrate models in periodic systems in quantum computing framework of Qiskit and Braket. Our results demonstrate that the hybrid quantum-classical approach can effectively model corrosion inhibitor interactions, providing a new computational tool for the screening and development of environmentally friendly corrosion inhibitors. The strong correlation between our computed binding energies and experimental inhibition efficiencies reported in literature \cite{SWATHI2022, ASHRAF2024} validates our computational approach and suggests its potential utility in future inhibitor design efforts. Yet, we would expect more accuracy by the quantum method if more orbitals are included in the calculations, more than the 5 orbitals used throughout the calculations due to the computational bottlenecks.

Our implementation is available at the project's GitHub repository \cite{inhibitQ2024}, which includes a comprehensive set of files and folders detailing the workflow we took and ensures easy implementation of our results; you can refer to the "phase2 submission" folder inside the repository "2024 inhibitQ". Where in each folder there are subfolders that detail the calculations inputs and results for each inhibitor and related algorithm used. For more details regarding hardware and other simulators experimentation, please refer to supplementary materials in hardware simulation. We also provided proof-of-concept for running the applied methodology in this article to run on real hardware (ion trap, superconducting).

\section*{Supplementary Information}
\label{supp_material}

\subsection*{S1. Computational Methods and Parameters}
\label{S1_computational_methods}
Table~\ref{tab:comp_details} provides a comprehensive overview of the computational parameters and methodologies employed in this work. The parameters were carefully chosen to balance computational efficiency with accuracy, particularly for the hybrid quantum-classical calculations of surface-adsorbate systems.

Our classical computational approach involved a two-step process of classical calculations then a hybrid quantum-classical approach: Preliminary geometry optimizations and fast electronic structure calculations were conducted using the Atomic Simulation Environment (ASE)\cite{ase-paper} and orb-d3-v2 model\cite{orbV2_2024}, which integrates Grimme's D3 dispersion corrections directly into the neural network potential. This choice was particularly important for our system containing substrate-adsorbate interactions, where dispersion forces play a crucial role. The orb models' native support for periodic systems made them especially suitable for our surface calculations, providing a balance between computational efficiency and accuracy in treating extended systems. For more details, please refer to their recent publication\cite{OrbPublication}. We want to add that our shortlisted inhibitors contain aromatic rings for $\pi$-$\pi$ stacking interactions and demonstrate strong electron transfer potential with aluminum surfaces\cite{Kaya2016}, that was based on the criteria we discussed in choosing inhibitors during phase1 submission.

We wanted to highlight that we have choosen a substrate of 4x4 to avoid the interactions between repeated cells in the xy directions between the images of the inhibitor molecule.

\begin{table*}
\caption{\label{tab:comp_details}Summary of Computational Methods and Parameters}
\begin{ruledtabular}
\begin{tabular}{ll}
Method/Component & Details \\
\hline
\multicolumn{2}{l}{\textbf{Classical Calculations}} \\
Geometry Optimization & ASE\cite{ase-paper} with orb-d3-v2 model\cite{orbV2_2024} \\
Dispersion Corrections & Grimme's D3 (integrated into neural network potential) \\
Surface Model & Al(111) 4×4 supercell \\
\hline
\multicolumn{2}{l}{\textbf{DFT Calculations}} \\
Functional & PBE\cite{Perdew1996} with GGA implementation\cite{Perdew1992} \\
Basis Set & DZVP-MOLOPT-GTH (double-zeta valence polarized) \\
Method & GPW\cite{Lippert1997}, plane-wave cutoff: 500 Ry, relative cutoff: 60 Ry \\
van der Waals & DFT-D3\cite{Grimme2010} with PBE reference functional \\
Vacuum Gap & 40 Å (z-direction) \\
Electronic Temperature & 1000 K (Fermi-Dirac distribution) \\
SCF Convergence & 1.0E-6 Ha, Broyden mixing ($\alpha$ = 0.1, $\beta$ = 1.5) \\
\hline
\multicolumn{2}{l}{\textbf{Active Space Parameters}} \\
Configuration & 2e, 5o (2 active electrons in 5 orbitals) \\
Selection Method & ActiveSpaceTransformer (Qiskit implementation), \\
& canonical orbital energy ordering selection\cite{Sayfutyarova2017} \\
\hline
\multicolumn{2}{l}{\textbf{Quantum Calculations}} \\
Primary Algorithm & ADAPT-VQE\cite{Grimsley2019} from Qiskit, \\
& StatefulAdaptVQE from qiskit-nature-cp2k\cite{qiskit_cp2k} \\
Qubit Mapping & Parity with two-qubit reduction\cite{Bravyi2017} \\
Convergence Criteria & Energy threshold: 1e-6 Hartree, gradient norm: 1e-4 \\
Classical Optimizer & SPSA\cite{Spall1992} (learning rate: 0.005, \\
& perturbation size: 0.05, max iterations: 1000) \\
\hline
\multicolumn{2}{l}{\textbf{Quantum Hardware \& Simulation}} \\
Simulators & Qiskit local and Aer\cite{qiskit_aer}, Braket local and SV1 \\
Hardware & IonQ Aria (via AWS Braket), IQM Garnet \\
Error Mitigation & Readout error mitigation\cite{Nachman2020} via Qiskit Runtime\cite{qiskit_zne}, \\
& Zero-noise extrapolation\cite{Temme2017} via Mitiq\cite{LaRose2022} \\
\end{tabular}
\end{ruledtabular}
\end{table*}

\subsection*{S2. Classical Calculations}
\begin{figure*}[ht]
    \centering
    \includegraphics[width=1.0\linewidth]{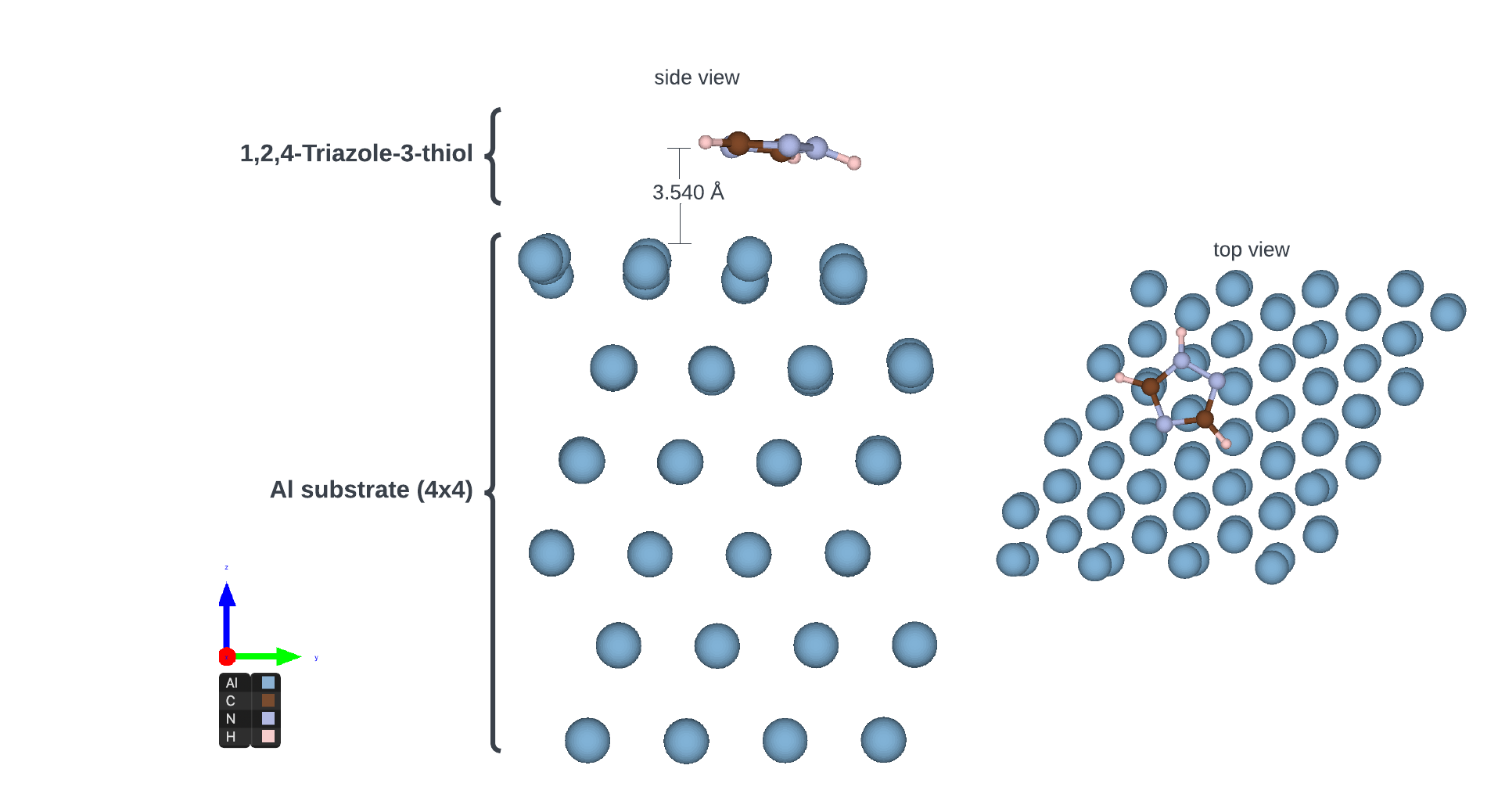}
    \caption{Side view and top view of geometry optimized supercell of 1,2,4-Triazole on top of Al substrate of size 4×4. The indicated binding distance was determined using the orb-d3-v2 ML potential model.}
    \label{fig:triazole_sc}
\end{figure*}

\begin{figure*}[ht]
    \centering
    \includegraphics[width=1.0\linewidth]{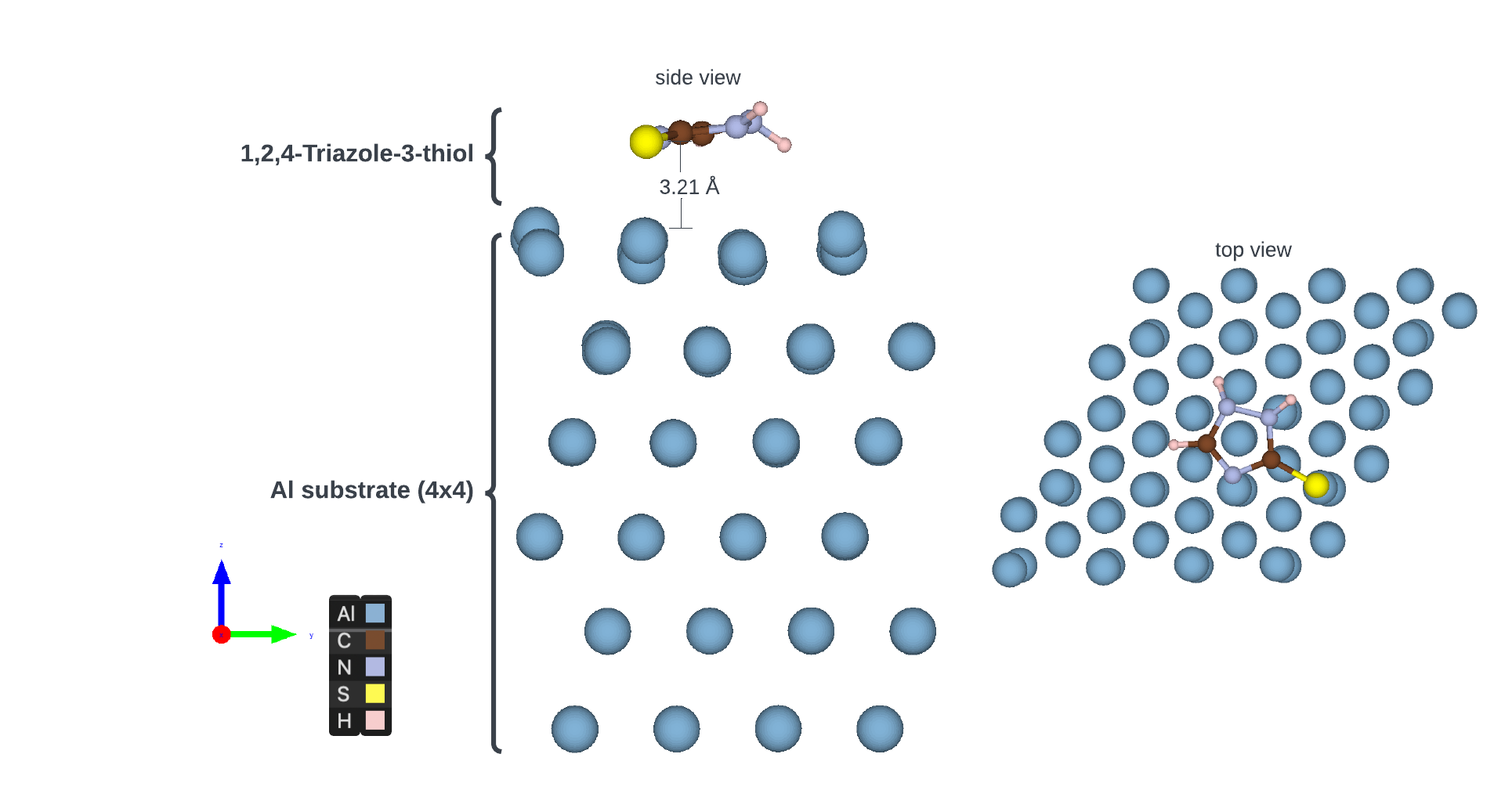}
    \caption{Side view and top view of geometry optimized supercell of 1,2,4-Triazole-3-thiol on top of Al substrate of size 4×4. The indicated binding distance was determined using the orb-d3-v2 ML potential model.}
    \label{fig:triazole_thiol_sc}
\end{figure*}

The optimized structures obtained from our classical calculations are shown in Fig.~\ref{fig:triazole_sc} and Fig.~\ref{fig:triazole_thiol_sc}. These structures demonstrate the successful application of our computational protocol, yielding binding geometries consistent with experimental observations~\cite{ASHRAF2024} and theoretical predictions~\cite{Winkler2016}.

We used the orb models' machine learning potentials to perform rapid geometry optimizations and binding energy calculations, comparing these classical computational results with quantum computing outcomes. Given that machine learning potentials are achieving accuracy comparable to DFT results, we relied on them to meet our time constraints. These ML potentials allowed us to efficiently explore multiple adsorption configurations and identify the most stable geometries for each inhibitor-surface system.

\subsection*{S3. Quantum Algorithm Benchmarking}
We conducted systematic benchmarking of various VQE implementations to validate our computational approach and establish optimal parameters for the surface-adsorbate calculations. To establish a robust computational workflow, we first benchmarked various algorithms of calculating the ground state energies using simpler systems using the LiH molecule. The benchmarking process consisted of three main components:

\subsubsection*{Standard VQE Algorithm Comparison}
Initial benchmarking compared three VQE implementations using LiH as a test system. Results showed that AdaptVQE demonstrated superior performance, reducing computational time by approximately 50\% while maintaining comparable accuracy (Table~\ref{tab:vqe_comparison}). This efficiency gain aligns with recent findings by Grimsley et al.~\cite{Grimsley2019} on the advantages of adaptive algorithms for electronic structure calculations.

\begin{table}[ht]
\caption{\label{tab:vqe_comparison}Comparison of VQE Algorithm Performance}
\begin{ruledtabular}
\begin{tabular}{lcc}
Algorithm & Runtime (s) & Ground State Energy (Ha) \\
\hline
VQE & 3654 & -7.8824 \\
StatefulVQE & 3134 & -7.8824 \\
AdaptVQE & 1552 & -7.8820 \\
\end{tabular}
\end{ruledtabular}
\end{table}

\subsubsection*{Adaptive VQE Variant Analysis}
Further comparison of different adaptive VQE implementations revealed that StatefulAdaptVQE exhibited exceptional performance, achieving a 5-6× speedup through its implementation of warm-starting techniques~\cite{Rossmannek2021} (Table~\ref{tab:adaptive_vqe}). The TetrisAdaptVQE implementation showed limited improvement opportunities, with only 12x the tetris feature applied in our test case.

\begin{table}[ht]
\caption{\label{tab:adaptive_vqe}Performance of Adaptive VQE Variants}
\begin{ruledtabular}
\begin{tabular}{lcc}
Algorithm & Runtime (s) & Ground State Energy (Ha) \\
\hline
AdaptVQE & 1258 & -7.8820 \\
StatefulAdaptVQE & 272 & -7.8802 \\
TetrisAdaptVQE & 1493 & -7.8820 \\
\end{tabular}
\end{ruledtabular}
\end{table}

\subsubsection*{Error Mitigation Assessment}
To evaluate quantum hardware implementation challenges, we conducted benchmarking studies using $\text{H}_2$ as a test system across different execution environments for AdaptVQE implementation in Qiskit (Table~\ref{tab:error_mitigation}). Our error mitigation strategy incorporated both readout error mitigation techniques~\cite{Nachman2020} and zero-noise extrapolation (ZNE)\cite{Temme2017, Kandala2019}, implemented through the Mitiq package\cite{LaRose2022}. The comparison between ideal simulation, noisy simulation using FakeVigo device, and FakeVigo with ZNE revealed substantial variations in ground state energies, from -1.1373 Ha in ideal conditions to 0.4083 Ha under noise. The application of ZNE significantly improved accuracy by recovering a ground state energy of -1.0305 Ha, demonstrating the crucial importance of error mitigation strategies in quantum hardware implementations. These results align with recent findings by Kandala et al.\cite{Kandala2019} and Temme et al.\cite{Temme2017}, and we further validated them through additional tests using local noisy simulators.

\begin{table}[ht]
\caption{\label{tab:error_mitigation}Impact of Error Mitigation on Ground State Energy Calculations}
\begin{ruledtabular}
\begin{tabular}{lc}
Environment & Ground State Energy (Ha) \\
\hline
Ideal Simulation & -1.1373 \\
FakeVigo (Noisy) & 0.4083 \\
FakeVigo + ZNE & -1.0305 \\
\end{tabular}
\end{ruledtabular}
\end{table}

\subsection*{S4. Active Space Analysis}
Our active space selection methodology combined the ActiveSpaceTransformer implementation from Qiskit with the embedding approach as implemented in CP2K code~\cite{cp2k2020} as detailed in their manual~\cite{cp2k_active_space}, utilizing the HARTREE-FOCK model to calculate active space interaction Hamiltonian. Following the recent framework by Battaglia et al.~\cite{Battaglia2024}, we employed the CANONICAL method for orbital selection, which orders orbitals based on their energy. This approach proved particularly effective for our surface-adsorbate systems, as the energy-ordered canonical orbitals naturally aligned with the orbitals around the Fermi level that are crucial for the adsorption process. The analysis revealed several significant electronic contributions to the corrosion inhibition mechanism, including the $\pi$-system of the triazole ring, the lone pair electrons on the sulfur atom (specifically for 1,2,4-Triazole-3-thiol), and the surface states at the Al(111) interface. These electronic features and their interactions align well with experimental observations by Winkler et al.~\cite{Winkler2016} and SWATHI et al.~\cite{SWATHI2022} regarding the mechanism of corrosion inhibition by triazole derivatives, validating our computational approach to active space selection.

\subsection*{S5. The adopted Hamiltonian}
\label{Hamiltonian}
The electronic Hamiltonian was constructed following the second-quantized formalism\cite{Rossmannek2021}:
\begin{equation}
\hat{H} = \sum_{pq} h_{pq} \hat{a}^\dagger_p\hat{a}q + \frac{1}{2}\sum{pqrs} g_{pqrs}\hat{a}^\dagger_p\hat{a}^\dagger_r\hat{a}s\hat{a}q
\end{equation}
where $h{pq}$ and $g{pqrs}$ represent the one- and two-electron integrals computed within the active space. The fermionic operators were mapped to qubit operators using the parity mapping with two-qubit reduction\cite{Bravyi2017}, which was chosen for its efficient handling of particle number conservation.

\subsection*{S6. Simulation SDKs and hardware}
\label{s4_simulation_hardware}
We made some effort in running our simulations on Braket SDK~\cite{AWS2024, Braket-github} through the integration tool qiskit-braket-provider~\cite{qiskit_braket}. We had to bugfix it for our special case of simulations (not adding any measurement instructions)~\cite{qiskit_braket_our}. Though at the time of writing simulations with braket local simulation SV1, IonQ Aria and IQM Garnet are still challenging, that is why our results presented in the results section focused on qiskit local simulations on HPC EC2 instances (Hpc6a, Hpc7a)~\cite{aws_hpc}.

For actual quantum computations, we used IonQ's Aria quantum processor and IQM's Garnet quantum processor, accessed through the AWS Braket platform. We managed to use it only for the benchmarking phase on small systems and not yet on our calculational cells. 

\subsection*{S8. Phase2 workflow and github link}
\begin{figure*}[ht]
    \centering
    \includegraphics[width=1.0\linewidth]{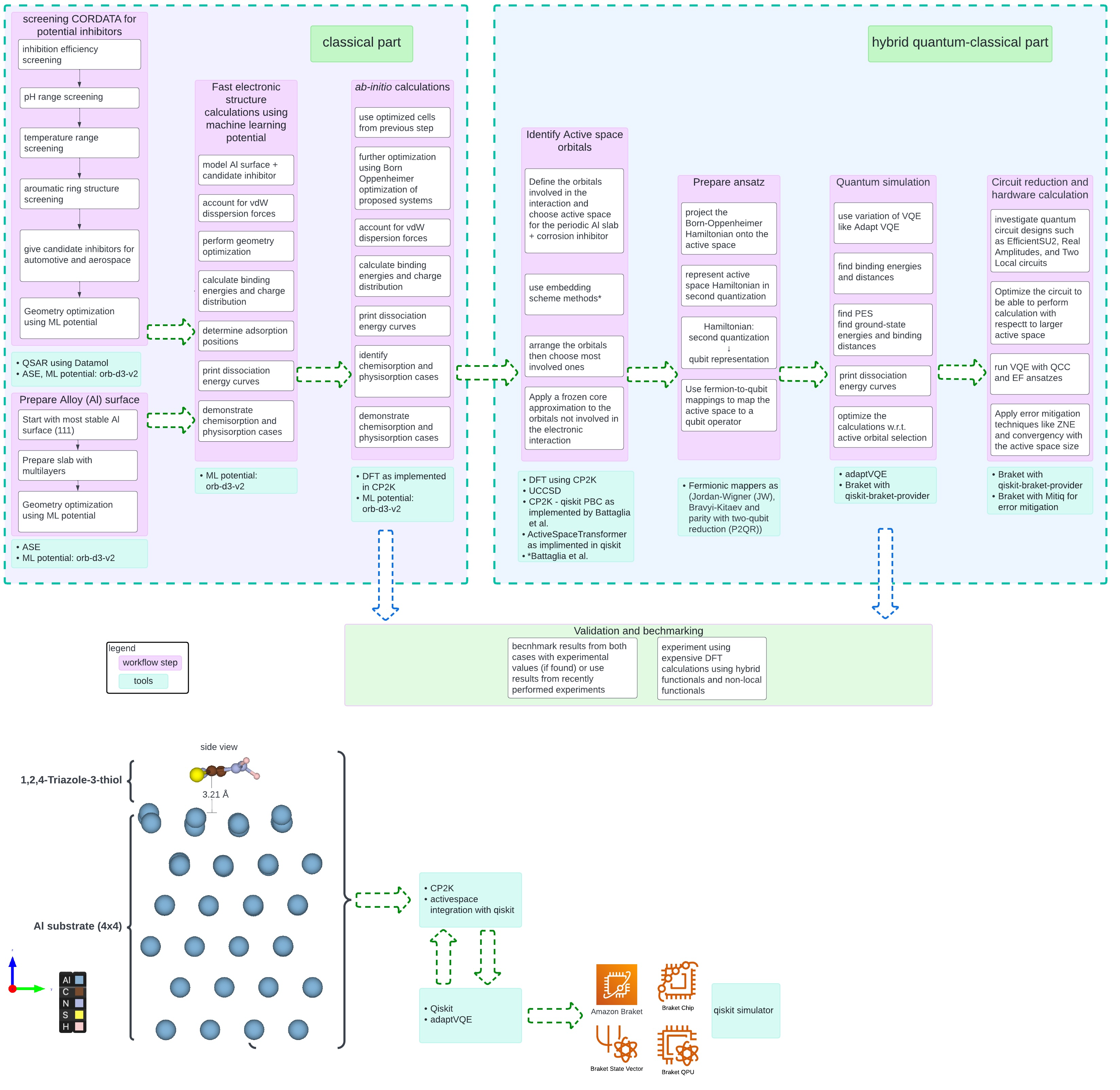}
    \caption{The workflow for phase2, illustrating the integration of classical and hybrid quantum-classical calculation steps to study the system.}
    \label{fig:workflow}
\end{figure*}

Fig.~\ref{fig:workflow} illustrates our integrated classical and quantum computational approach that we introduced during phase 1 and edited a bit in phase 2, yet we believe that the main core functionalities are still there. The workflow emphasizes the steps from classical geometry optimization through quantum electronic structure calculations and algorithm application on the chosen systems. The github link for all details can be found here (\url{https://github.com/MarcMaussner/2024_inhibitQ/tree/main/phase2_submission}) and the repo itself is here (\url{https://github.com/MarcMaussner/2024_inhibitQ/tree/main/phase2_submission})

\subsection*{Acknowledgements}
This work is a result of our participation in the mobility challenge organized by Airbus, BMW and AWS Braket. We want to thank the resonance team for their efforts in making the administrative work and communication easy and clear. We are grateful for the feedback sessions with Airbus and BMW experts. We thank all of them for the discussions during the expert sessions. We would like to extend our appreciation to the Braket team, particularly for their warm welcome and availability in connecting with us and following up to solve challenges. Lastly, we would like to thank the IonQ team for their willingness to help us gain the best knowledge of their hardware offering.

\nocite{*}

\bibliography{apssamp}

\end{document}